# Properties of Hydrogen Bonded Networks in Ethanol-water Liquid Mixtures as a Function of Temperature: Diffraction Experiments and Computer Simulations

*Szilvia Pothoczki\*[1], Ildikó Pethes[1], László Pusztai[1,2], László Temleitner[1], Koji Ohara[4], and Imre Bakó\*[3]*

[1]Wigner Research Centre for Physics, H-1121 Budapest, Konkoly Thege M. út 29-33., Hungary

[2]International Research Organization for Advanced Science and Technology (IROAST), Kumamoto University, 2-39-1 Kurokami, Chuo-ku, Kumamoto, 860-8555, Japan

[3]Research Centre for Natural Sciences, H-1117 Budapest, Magyar tudósok körútja 2., Hungary

[4]Diffraction and Scattering Division, JASRI, Spring-8, 1-1-1, Kouto, Sayo-cho, Sayo-gun, Hyogo 679-5198, Japan

**Abstract**

New X-ray and neutron diffraction experiments have been performed on ethanol-water mixtures as a function of decreasing temperature, so that such diffraction data are now available over the entire composition range. Extensive molecular dynamics simulations show that the all-atom interatomic potentials applied are adequate for gaining insight of the hydrogen bonded network structure, as well as of its changes on cooling. Various tools have been exploited for revealing details concerning hydrogen bonding, like determining H-bond acceptor and donor sites, calculating cluster size distributions and cluster topologies, as well as computing the Laplace spectra and fractal dimensions of the networks. It is found that 5-membered hydrogen bonded cycles are dominant up to an ethanol content of 70% at room temperature, above which concentration ring structures nearly disappear. Percolation has been given special attention, so that it could be shown that at low temperature, close to the freezing point even the mixture with 90% ethanol possesses a 3D percolating network. Moreover, the water sub-network also percolates even at room temperature, with a percolation transition occurring around 50% ethanol.

# INTRODUCTION

Physico-chemical properties of water-ethanol solutions have been among the most extensively studied subjects in the field of molecular liquids over the past few decades[1-17], due to their high biological and chemical significance. Even though they are composed of two simple molecules, the behavior of their hydrogen bonded network structures can be very complex, due the competition between hydrophobic and hydrophilic interactions.[18-25] The characteristics of these networks can be greatly influenced by the concentration. Usually, three regions of the composition range are distinguished qualitatively: the water rich, the medium or transition, and the alcohol rich regions.

Most thermodynamic properties, such as excess enthalpy, isentropic compressibility and entropy, show either maxima or minima in the low alcohol concentration region, the molar ratio of ethanol $x_{eth}<0.2$.[26-28] Differential scanning calorimetry, NMR and IR spectroscopic studies suggested a transition point around $x_{eth} = 0.12$, while additional transition points were found at $x_{eth} = 0.65$ and $0.85$.[29-31] Concerning the intermediate region around $x_{eth}=0.5$, a maximum was observed by the Kirkwood-Buff integral theory, which suggests water-water aggregation.[32-35] Also, a maximum of the concentration fluctuations was found in the same region, at $x_{eth}=0.4$, by small angle X-ray scattering.[36]

Quite recently we studied structural changes in ethanol-water mixtures as a function of temperature in the water rich region (up to $x_{eth}=0.3$).[23-24] There we focused mainly on the cyclic entities. We found that the number of hydrogen bonded rings has increased with lowering the temperature, and that five-fold rings were in majority, especially at $x_{eth}>0.1$ ethanol concentrations.

In the present study, we extend both X-ray diffraction measurements and molecular dynamics simulations to investigating ethanol-water mixtures down to their freezing points, over the entire ethanol concentration range. Furthermore, new neutron diffraction experiments have been performed in the water rich region (up to $x_{eth}$=0.3). These neutron data fit nicely in the present line of investigation and support our earlier findings. The main goal here was to provide a complete picture of the behavior of the hydrogen bonded network over the entire composition range in ethanol-water mixtures, between room temperature and the freezing point. In order to identify the existence and the location of the percolation threshold, we monitor the changes of the number of molecules acting as donor or acceptor, cluster size distributions, cyclic and non-cyclic properties, and the Laplace spectra of the H-bonded network.

## METHODS

**X-ray and neutron diffraction experiments**

Series of samples of ethanol-water mixtures have been prepared with natural isotopic abundances for synchrotron X-ray, and with fully deuterated forms of both compounds for neutron diffraction experiments.

Synchrotron experiments were performed at the BL04B2[37] high energy X-ray diffraction beamline of the Japan Synchrotron Radiation Research Institute (SPring-8, Hyogo, Japan). Diffraction patterns could be obtained over a scattering variable, $Q$, range between 0.16 and 16 Å$^{-1}$, for samples with alcohol contents of 40, 50, 60, 70, 80, 85, 90 and 100 mol % of ethanol. Diffraction patterns have been recorded starting from room temperature and cooling down to freezing point for each composition.

Neutron diffraction measurements have been carried out at the 7C2 diffractometer of Laboratoire Léon-Brillouin[38]. Details of the experimental setup, the applied ancillary equipment and data correction procedure were already reported[25] for methanol-water samples measured under the same conditions. For both neutron and X-ray raw experimental data, standard procedures[38,39] have been applied during data treatment.

All temperature and composition points visited by the new X-ray and neutron diffraction experiments are displayed together with the phase-diagram of ethanol-water mixtures in Fig.1. Total scattering structure factors (TSSF) obtained from the new experimental data are shown in Fig. 2 and Fig. S1.

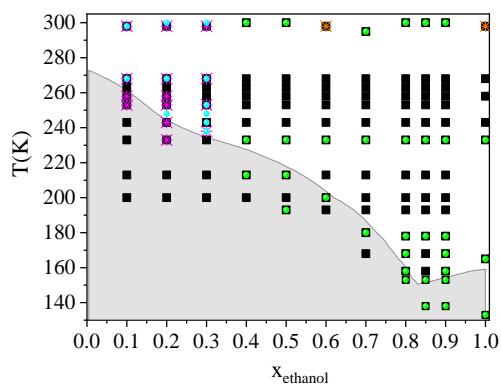

**Figure 1.** Phase diagram of ethanol-water mixtures[40]. Gray area: solid state; white area: liquid state (as determined experimentally). Black solid squares: present MD simulations; green solid circles: new X-ray diffraction data sets; light blue solid circles: new neutron diffraction data sets; orange crosses: X-ray diffraction data sets from Ref. 19.; magenta crosses: X-ray diffraction data sets from Ref. 21.

**Molecular Dynamics (MD) Simulations**

Molecular Dynamics (MD) simulations were carried out by using the GROMACS software[41] (version 2018.2). The Newtonian equations of motions were integrated by the leapfrog algorithm, using a time step of 2 fs. The particle-mesh Ewald algorithm was used for handling long-range electrostatic forces.[42-43] The cut-off radius for non-bonded interactions was set to 1.1 nm. For ethanol molecules, the all-atom optimized potentials for liquid simulations (OPLS-AA)[44] force field was used. Bond lengths were kept fixed by the LINCS algorithm[45]. Parameters of atom types and atomic charges can be found in Table S1. Based on results of our earlier study[22], the TIP4P/2005[46] water model was applied, as handled by the SETTLE algorithm[47].

For each composition, 3000 molecules (with respect to compositions and densities) were placed in a cubic box, with periodic boundary conditions. Box lengths, together with corresponding bulk densities can be found in Table S2. All MD models studied are identified in Fig.1. Table S3 contains the various phases of the MD simulations.

**RESULT AND DISCUSSION**

**Total scattering structure factors**

As typical examples, total scattering structure factors obtained from measured X-ray diffraction signals, for 40 mol% and 50 mol% aqueous solutions of ethanol, as a function of temperature, are drawn in Fig. 2a. Similarly, Fig. 2b shows TSSF-s from neutron diffraction for $x_{eth}=0.3$. Calculated TSSF-s are also presented in Figure 2. Additional measured TSSF-s, together with the corresponding calculated TSSF-s, can be found in Fig. S1.

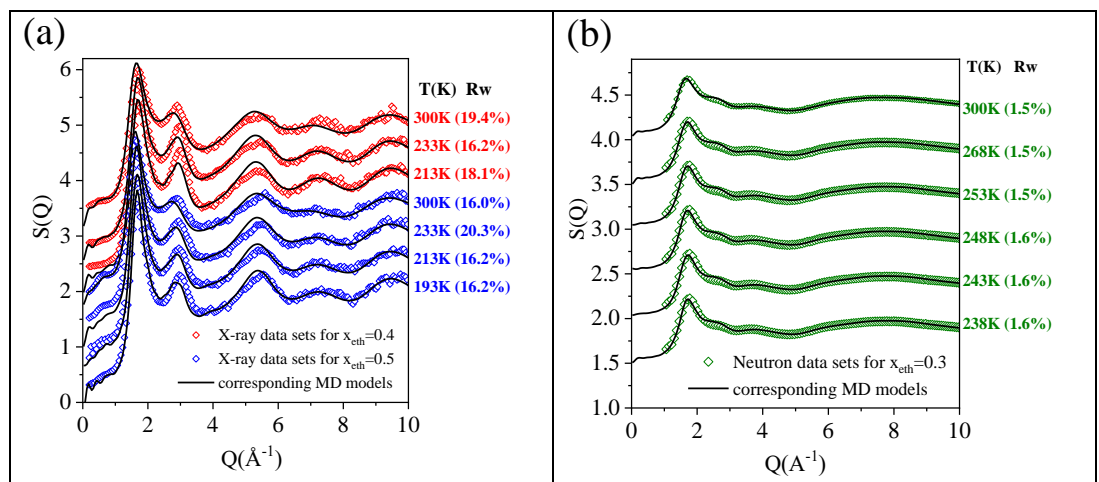

**Figure 2.** Measured and calculated TSSF's a) for X-ray diffraction; b) for neutron diffraction.

Agreement between calculated and measured TSSF-s for neutron diffraction appears to be almost perfect. In the case of X-ray diffraction, apparent differences can be observed, mostly around the second maximum. $R_w$ factors were calculated to characterize differences between MD simulated [$F^S(Q)$] (averaged over many time frames) and experimental structure factors [$F^E(Q)$] quantitatively, thus providing a kind of goodness-of-fit (c.f. Supp. Info.). Note that values of $R_w$ for the two different experimental methods are not to be compared, due to the different data treatment procedures. It can be stated that MD models are appropriate for further analyses.

We note here that detailed analyses of partial radial distribution functions (prdf's) are not within the scope of the present work, however all prdf's related to H-bonding properties appear in Figs. S2-S8.

**H-bond acceptors and donors**

The calculated average hydrogen bond numbers for the entire mixture and for the ethanol subsystem can be found in Figs. S9-S11. The H-bond definition applied is presented also in the Supp. Info. All of the following analyses (together with the identification of cyclic and non-cyclic entities) were performed by using our in-house computer code.[48]

Molecules participating in H-bonds can be classified into two groups according to their roles as proton acceptors or donors. Each molecule may have a certain number of donor sites ($n_D$) and a certain number of acceptor sites ($n_A$), and thus can be characterized by the '$n_DD:n_AA$' combination. For example, '1D:2A' denotes a molecule which acts as a donor of 1 H-bond and accepts 2 H-bonds. The sum of $n_D$ and $n_A$ for a given molecule provides the number of H-bonds ($n_{HB}$) of that molecule. (c.f. Fig. S9.)

The most populated fractions for ethanol molecules are '1D:1A', '1D:2A' and the sum of '0D:1A' and '1D:0A' (Fig. 3a), while for water molecules they are '1D:2A', '2D:1A' and '2D:2A' (Fig. 3b). These groups altogether contain 80% of all H-bonds at room temperature and 90% of all H-bonds at low temperatures.

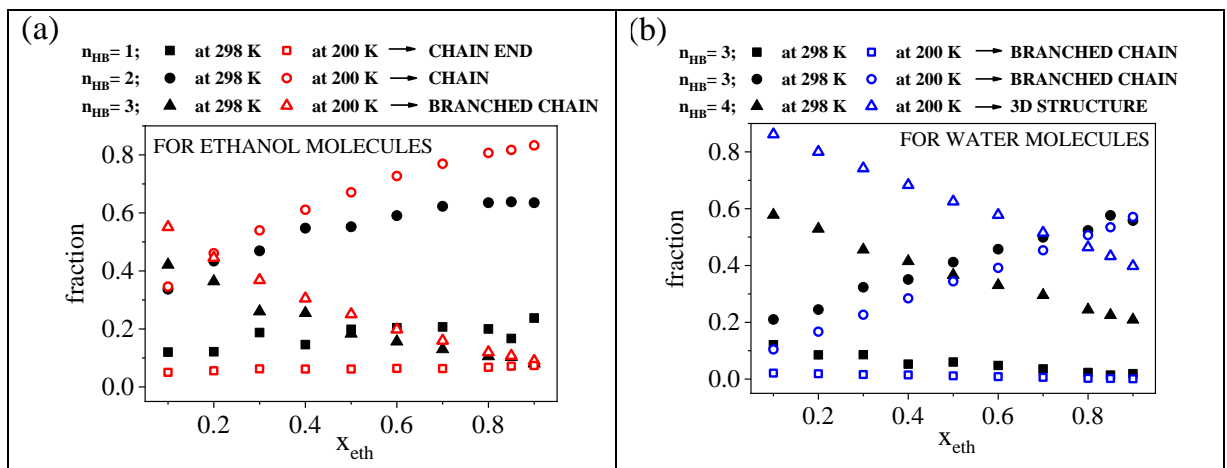

**Figure 3.** Donor and acceptor sites a) for ethanol molecules and b) for water molecules as a function of ethanol concentration.

Concerning ethanol molecules, the occurrence of the '1D:1A' combination is above 50% over almost the entire concentration range, independently of the temperature. This group corresponds to chain-like arrangements, which become more preferred with increasing ethanol content and at lower temperatures. It is remarkable that in the water-rich region the '1D:2A' combination has the same as ($x_{eth}$=0.2), or even a slightly higher occurrence ($x_{eth}$=0.1) than that of '1D:1A'. With decreasing water content, the occurrence of '1D:2A' decreases. Also, this group is more dominant at 200 K.

Water molecules most often behave according to the '2D:2A' scheme. The occurrence of this arrangement significantly increases with decreasing temperature, as well as with increasing water content. On the other hand, the fractions of '1D:2A' and '2D:1A' combinations increase as temperature increases. There is a well-defined asymmetry between these two ('1D:2A' and '2D:1A') types of water molecules in terms of their populations, which difference becomes more pronounced with increasing ethanol concentration. The fractions of '1A:1D' for ethanol molecules, '1D:2A' for ethanol molecules, '2D:1A' for water molecules, '2D:2A' for water molecules as a function of temperature can be found in Fig. S12. Furthermore, calculated H-bond number excess parameter is shown in Fig. S13.

**Clustering and percolation**

Two molecules are regarded as members of a cluster, according to the definitions introduced by Geiger et al.[49], if they are connected by a chain of hydrogen bonds. Concerning the pure components of the mixtures studied here, water molecules are forming a three dimensional percolating hydrogen bonding network[46,49], whereas in pure ethanol only chain (or branched chain) structures can be detected.[50,51]

There are several descriptors connected to the properties of networks that can be used for the determination of the percolation transition. This work focuses on the cluster size distribution ($P(n_c)$). However, very similar conclusions may be drawn from scrutinizing several other parameters such as the average largest cluster size (C1), average second largest cluster size (C2), and the fractal dimension of the largest cluster ($f_d$). More detailed discussion is provided in the Supporting Information, below Fig. S14.

Cluster size distributions are shown in Figure 4. The system is percolated when the number of molecules in the largest cluster is in the order of the system size. For random percolation on a 3D cubic lattice, the cluster size distribution can be given by $P(n_c)=n_c^{-2.19}$ ($n_c$ is the number of molecules in a given cluster).[52,53] Percolation transition can be ascertained by comparing the calculated cluster size distribution function of the present system with that obtained for the random one. At each temperature up to $x_{eth}=0.85$ (Fig. 4 and Fig. S15a) a well-defined contribution can be found at large cluster size values, signaling percolation. Systems with $x_{eth}=0.9$ (Fig. 4b) show the same behavior at lower temperatures, but this signature disappears at room temperature. This suggests that in the latter case the system is close to the percolation threshold, which can be expected between 0.9 and 1.0 ethanol molar fraction. Ethanol molecules in the pure liquid compose non-percolated assemblies (Fig. S15b).[2,8,20]

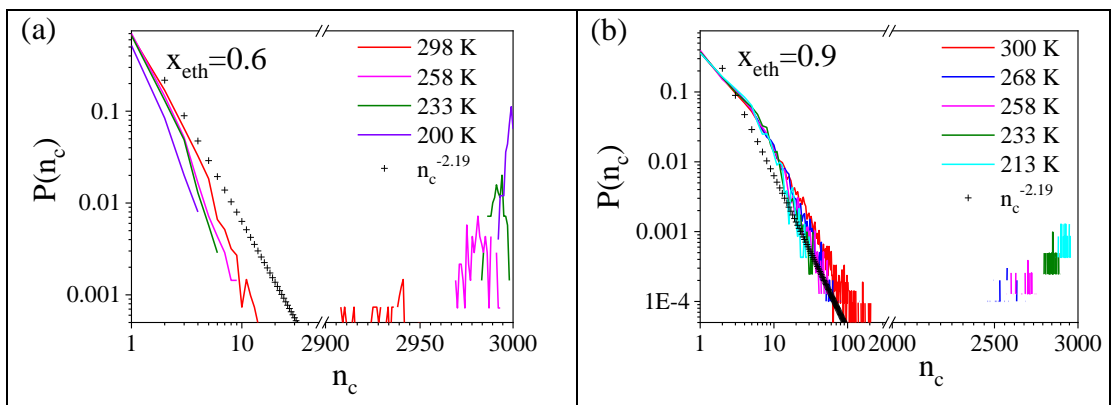

**Figure 4**. Cluster size distributions from the room temperature to the lowest studied temperature a) for $x_{eth}=0.6$, b) for $x_{eth}=0.9$.

The role of water molecules was then analyzed separately. All of the four quantities mentioned above for characterizing the percolation transition were calculated, taking into account only H-bonds between water molecules. Fig. 5 shows one representation. The average largest cluster size divided by the total number of water molecules drops down below 0.5, which the sign of the percolation transition, between $x_{eth}=0.4$ and $x_{eth}=0.5$ at 300 K, and between $x_{eth}=0.5$ and $x_{eth}=0.6$ at 200 K, respectively. Similar values were found for percolation in formamide-water[54] and glycerol-water mixtures[55]. However, in those cases both of the constituents (not only water molecules) form 3D percolating H-bonded networks in the liquid state. Here, in contrast, independently of the concentration, ethanol molecules form only short chain-like structures, but not large percolated networks. Typical hydrogen bond network topologies for the largest cluster at the composition of $x_{eth}=0.4$, 0.7 and 0.9 are shown in Fig. S16.

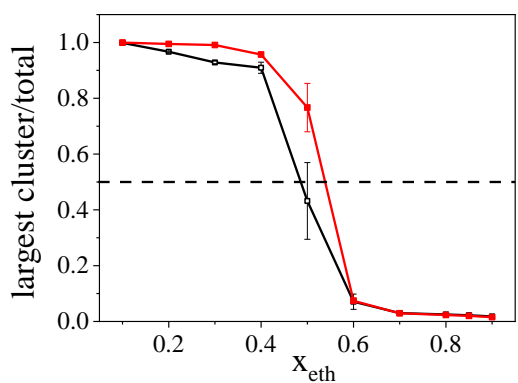

**Figure 5.** The average largest cluster size divided by the total number of water molecules, as a function of the ethanol concentration. Black open square symbols: 300 K, red solid square symbols: 200 K.

**Rings and chains**

Hydrogen bonded clusters may contain non-cyclic, chain-like and cyclic, "closed into themselves" entities (c.f. Fig. S16). The number of cyclic entities ($N_{cycl}$), the number of molecules ($N_{noncycl}$) that are not members of any ring ($n_c < 10$) and the cyclic size distribution ($n_r$) were calculated using the algorithms developed by Chihaia et al.[56]

Figure 6 summarizes the numbers of cyclic and non-cyclic entities as a function of ethanol concentration and temperature. The number of cycles decreases significantly with increasing ethanol content. As a result, in the ethanol rich region (above 70 mol%) mostly non-cyclic entities are present. Both the $N_{noncycl}$ and $N_{cycl}$ entities show a strong temperature dependence up to around $x_{eth}=0.80-0.85$. This effect appears to be more pronounced for non-cyclic entities. At the highest ethanol concentrations, where most of the molecules are arranged in chains, the number of chains formed is independent of temperature.

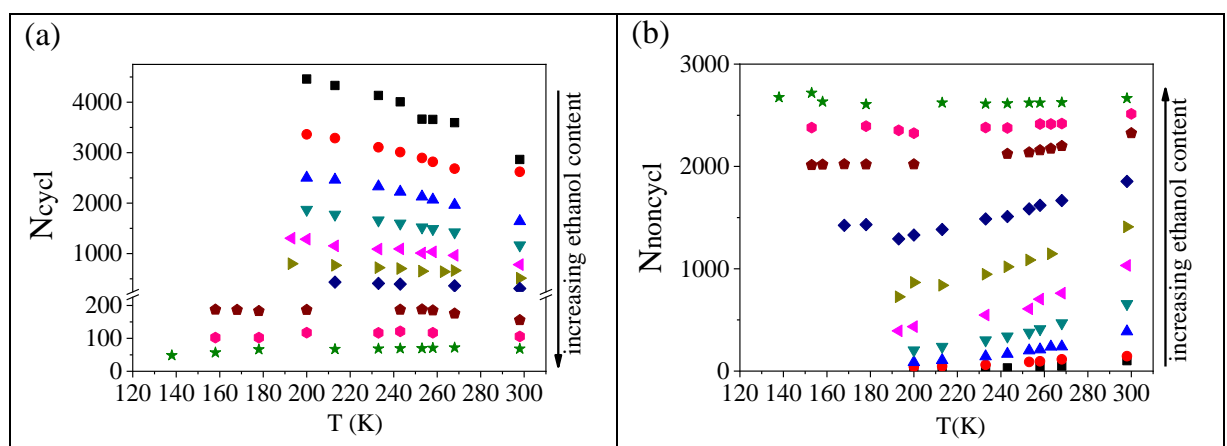

**Figure 6.** a) Number of cyclic entities, and b) number of non-cyclic entities as a function of ethanol concentration and temperature. Black squares: $x_{eth}=0.1$; red circles: $x_{eth}=0.2$; blue up triangles: $x_{eth}=0.3$; dark cyan down triangles: $x_{eth}=0.4$; magenta left triangles: $x_{eth}=0.5$; dark

yellow right triangles: $x_{eth}$=0.6; navy diamonds: $x_{eth}$=0.7; dark red pentagons: $x_{eth}$=0.8; dark magenta hexagons: $x_{eth}$=0.85; green stars: $x_{eth}$=0.9.

It has already been demonstrated that in pure water, molecules prefer to form six-membered rings at room temperature, which behaviour becomes more pronounced during cooling.[24] This statement remains true in ethanol-water mixtures (Fig. 7), as well, as long as the ethanol molar ratio stays around 0.1, whereas for $x_{eth}$=0.2 and 0.3, 5-membered rings become dominant[22]. Regardless of the ethanol concentration, there are always more rings at low temperatures[22,24]. Focusing now on mixtures with ethanol contents higher than 30 mol%, 5-membered rings take the leading role up to a concentration somewhere between $x_{eth}$=0.7 and 0.8, where the number of rings (per particle configuration) falls below 100. These tendencies are more pronounced at lower temperatures.

Note that for the sake of comparison, results for the region between $x_{eth}$=0.1 and 0.3 are also presented in Figure 7, although detailed discussions of ring size distributions for the water rich region can be found in Ref. 22. The new element here is that the corresponding curves are consistent also with our fresh neutron diffraction data (cf. Fig. 2b).

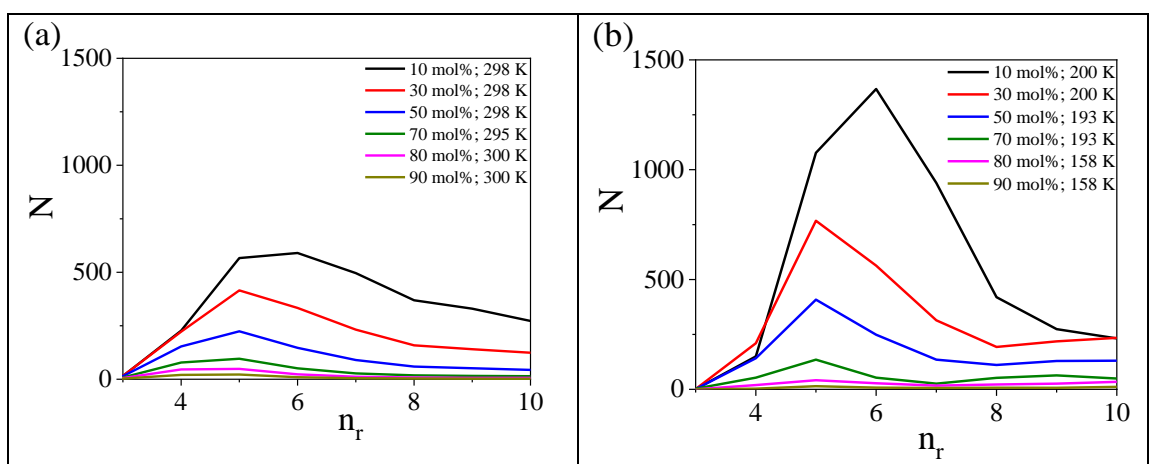

**Figure 7.** Ring size distributions as a function of $x_{eth}$. a) At room T; b) at the studied lowest T.

**Spectral properties of H-bonded networks**

It has already been shown that Laplace spectra[57-67] of H-bonded networks is a good topological indicator for monitoring the percolation transition in liquids[68]. Several authors have studied the relationship between the eigenvector corresponding to the second smallest eigenvalue ($\lambda_2$) and the graph structure; well documented reviews can be found in the literature.[58,62,65] More details are placed in Supp. Info.

Figure 8 provides Laplace spectra of ethanol-water mixtures as a function of concentration at room temperature. The low $\lambda$ values (up to 0.3) are enlarged at the bottom. Spectra of the pure constituents can be found in Ref. 68. According to the topology of the H-bonded network, two cases can be distinguished in connection with Laplace-spectra: (1) For liquids whose molecules are forming a 3D percolated network, a well-defined gap can be detected at low eigenvalues. (2) For systems without extended 3D network structure, where molecules link to each other so that to construct (branched) chains, several well-defined peaks ($\lambda = 0.5, 1, 1.5, 2…$etc.) show up, without any recognizable gap at low eigenvalues. Pure water falls into the first category, while pure liquid ethanol belongs to the second one.[68]

Concerning the mixtures studied here, the existence of the gap mentioned above depends on composition. Up to an ethanol content of 95 % a well-defined gap can be found. However, above 95% ethanol content this gap disappears. Concerning the H-bond network, this means that there is a limiting alcohol concentration beyond which the presence of chains of molecules is dominant. In this concentration region, percolated networks cannot be detected. That given concentration when percolation vanishes can be considered as a percolation threshold. At concentrations lower than this limiting value, water-like 3D percolated networks are formed. At lower temperatures no percolation threshold could be found, all systems have 3D network structures (cf. Fig. 4b).

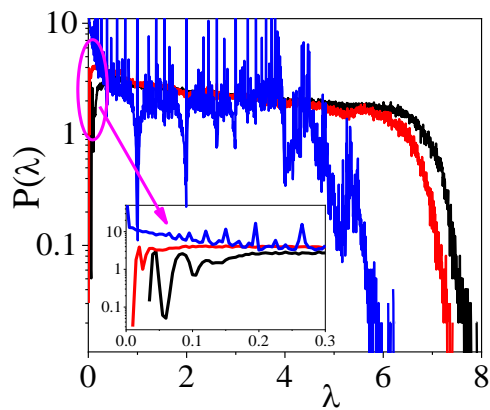

**Figure 8.** The Laplace spectra of ethanol-water mixtures as a function of concentration at room temperature. Black line: $x_{eth}=0.1$; red line: $x_{eth}=0.4$; blue line: $x_{eth}=0.95$.

**SUMMARY AND CONCLUSIONS**

X-ray and neutron diffraction measurements have been conducted on ethanol-water mixtures, as a function of temperature, down to the freezing points of the liquids. As a result of the new experiments, temperature dependent X-ray structure factors are now available for the entire composition range.

For interpreting experimental data, series of molecular dynamics simulations have been performed for ethanol-water mixtures with ethanol contents between 10 mol% and 90 mol%. Temperature has been varied between room temperature and the freezing point of the actual mixture. With the aim of evaluating the applied force fields, MD models have been compared to new X-ray diffraction data over the entire composition range, as well as to new neutron diffraction experiments over the water rich region. It has been established that the combination of OPLS-AA (ethanol) and TIP4P/2005 (water) potentials have reproduced individual experimental data sets, as well as their temperature dependence, with a more than satisfactory accuracy. It may therefore be justified that the MD models are used for characterizing hydrogen bonded networks that form in ethanol-water mixtures.

When H-bond acceptor and donor roles of water molecules are taken into account, the occurrence of the '2D:2A' combination increases linearly at every concentration with decreasing temperature.

The percolation threshold and its variation with temperature has been estimated via various approaches: we found that even at the highest alcohol concentration, the entire system percolates at low temperatures. The percolation transition for the water subsystems was found to be a 3D percolation transition that occurs between $x_{eth}=0.4$ and $x_{eth}=0.5$ at 300 K, and between $x_{eth}=0.5$ and $x_{eth}=0.6$ at 200 K, respectively.

Concerning the topology of H-bonded assemblies, in mixtures with ethanol contents higher than 30 mol%, 5-membered ring take the leading role up to the $x_{eth}=0.7$ and 0.8, where the number of rings falls dramatically. This tendency is most pronounced at low temperatures.

**Supporting Information**

Measured and calculated total scattering structure factors for X-ray diffraction for the ethanol-water mixtures as a function of temperature (Fig. S1); Lennard-Jones parameters and partial charges for the atom types of ethanol used in the MD simulations (Table S1); Steps of Molecular Dynamics simulation at each studied temperature (Table S2); Box lengths (nm), corresponding bulk densities (g/cm$^3$) for each simulated system (Table S3); Selected partial radial distribution functions for the mixture with 40 mol%, 50 mol%, 60 mol%, 70 mol%, 80 mol%, 85 mol% and 90 mol% ethanol as a function of temperature (Figs. S2-S8); Average H-bond numbers considering each molecule, regardless of their types, together with the case when considering water–water H-bonds only (Fig. S9); Average H-bond number for ethanol-ethanol subsystem (Fig. S10); Average H-bond numbers considering connections of water molecules only, as well as considering connections of ethanol molecules only (Fig. S11); Fraction of donor and acceptor sites as a function of temperature (Fig. S12); H-bond number excess parameter (Fig. S13); The average largest cluster size (C1) and average second largest cluster size (C2) as a function of ethanol concentration and temperatures in ethanol-water mixtures (Fig. S14); Cluster size distributions from the room temperature to the lowest studied temperature a) for $x_{eth}$=0.85, b) for pure ethanol (Fig. S15); Typical hydrogen bonded network topologies in water-ethanol mixtures at concentrations $x_{eth}$ = 0.40 (left), 0.70 (middle), and 0.90 (right) (Fig. S16); Values of the inequality calculated by Equation 4. Black open squares: left side of Eq. 4 at 298 K; black solid squares: right side of Eq. 4 at 298 K; red open circles: left side of Eq. 4 at 233 K; red solid circles: right side of Eq. 4. at 233 K (Fig. S17).

AUTHOR INFORMATION

**Corresponding Authors**

**Szilvia Pothoczki** – Wigner Research Centre for Physics, H-1121 Budapest, Konkoly-Thege M. út 29-33., Hungary; Email: pothoczki.szilvia@wigner.hu

**Imre Bakó** – Research Centre for Natural Sciences, H-1117 Budapest, Magyar tudósok körútja 2., Hungary; Email: bako.imre@ttk.hu

**Authors**

**Ildikó Pethes** – Wigner Research Centre for Physics, H-1121 Budapest, Konkoly-Thege M. út 29-33., Hungary;

**László Pusztai** – Wigner Research Centre for Physics, H-1121 Budapest, Konkoly-Thege M. út 29-33., Hungary; International Research Organization for Advanced Science and Technology (IROAST), Kumamoto University, 2-39-1 Kurokami, Chuo-ku, Kumamoto, 860-8555, Japan

**László Temleitner** – Wigner Research Centre for Physics, H-1121 Budapest, Konkoly -hege M. út 29-33., Hungary;

**Koji Ohara** – Diffraction and Scattering Division, JASRI, Spring-8, 1-1-1, Kouto, Sayo-cho, Sayo-gun, Hyogo 679-5198, Japan

**Acknowledgement**

The authors are grateful to the National Research, Development and Innovation Office (NRDIO (NKFIH), Hungary) for financial support via grants Nos. KH 130425, 124885 and FK 128656. Synchrotron radiation experiments were performed at the BL04B2 beamline of SPring-8 with the approval of the Japan Synchrotron Radiation Research Institute (JASRI) (Proposal Nos. 2017B1246 and 2018A1132). Neutron diffraction measurements were carried out on the 7C2 diffractometer at the Laboratoire Léon Brillouin (LLB), under Proposal id. 378/2017. Sz. Pothoczki and L. Temleitner acknowledge that this project was supported by the János Bolyai Research Scholarship of the Hungarian Academy of Sciences. The authors thank J. Darpentigny (LLB, France) for the kind support during neutron diffraction experiment. Valuable assistance from Ms. A. Szuja (Centre for Energy Research, Hungary) is gratefully acknowledged for the careful preparation of mixtures.

Supporting Information

for

*Properties of hydrogen bonded network in ethanol-water liquid mixtures as a function of temperature: diffraction experiments and computer simulations*

*Szilvia Pothoczki*[1], *Ildikó Pethes*[1], *László Pusztai*[1,2], *László Temleitner*[1], *Koji Ohara*[4], *and Imre Bakó*[3]*

[1]Wigner Research Centre for Physics, H-1121 Budapest, Konkoly Thege M. út 29-33., Hungary

[2]International Research Organization for Advanced Science and Technology (IROAST), Kumamoto University, 2-39-1 Kurokami, Chuo-ku, Kumamoto, 860-8555, Japan

[3]Research Centre for Natural Sciences, H-1117 Budapest, Magyar tudósok körútja 2., Hungary

[4]Diffraction and Scattering Division, JASRI, Spring-8, 1-1-1, Kouto, Sayo-cho, Sayo-gun, Hyogo 679-5198, Japan

*E-mail: pothoczki.szilvia@wigner.hu; Phone: +36 1 392 2222/ext. 1469

*E-mail: bako.imre@ttk.hu; Phone: +36 1 382 6981

**Figure S1:** Measured and calculated total scattering structure factors (TSSF's) for the ethanol-water mixtures as a function of temperature. a) X-ray diffraction TSSF's for $x_{eth}$=0.8 and 0.85; b) X-ray diffraction TSSF's for $x_{eth}$=0.9 and 1.0; c) X-ray diffraction TSSF's for $x_{eth}$=0.6 and 0.7; d) neutron diffraction TSSF's for $x_{eth}$=0.1 and 0.2.

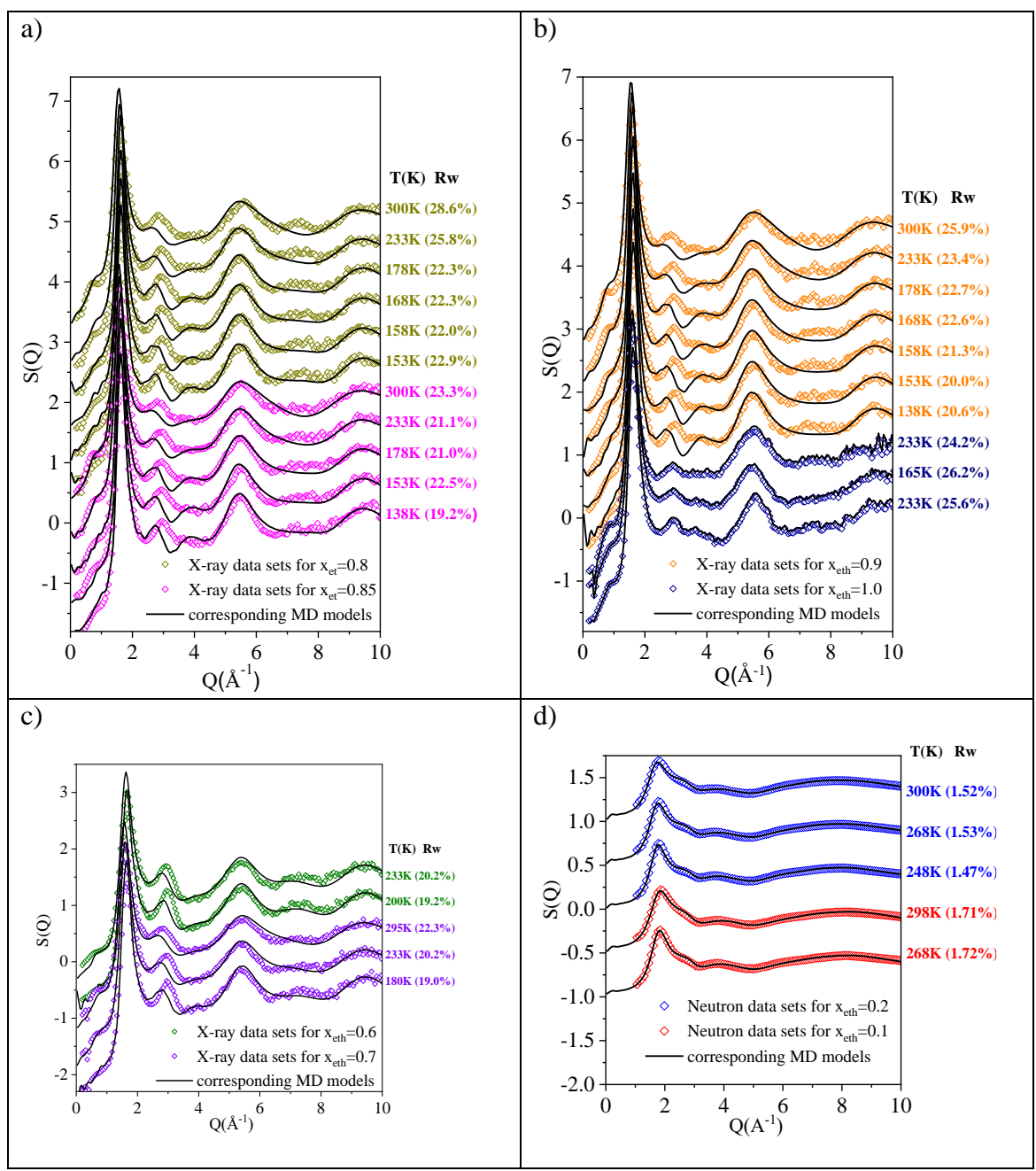

**Table S1:** Lennard-Jones parameters and partial charges for the atom types of ethanol used in the MD simulations.

| Atom types | σ (Å) | ε (kJ/mol) | q (e) |
|---|---|---|---|
| C (H$_3$) | 3.5 | 0.276144 | -0.18 |
| C (H$_2$) | 3.5 | 0.276144 | 0.145 |
| H | 2.5 | 0.12552 | 0.06 |
| O | 3.12 | 0.71128 | -0.683 |
| OH | 0 | 0 | 0.418 |

**Table S2**: The first rows: box lengths (nm), the second rows: the corresponding bulk densities (g/cm$^3$) for each simulated system. One star symbol: at 298; two stars symbol: at 295 K; without symbol: at 300 K.

| x$_{eth}$ | 0.1 | 0.2 | 0.3 | 0.4 | 0.5 | 0.6 | 0.7 | 0.8 | 0.85 | 0.9 | 1.0 |
|---|---|---|---|---|---|---|---|---|---|---|---|
| 298 K | 4.750* | 4.998* | 5.243* | 5.479 | 5.685 | 5.865* | 6.052** | 6.247 | 6.354 | 6.441 | 6.608* |
|       | 0.9678 | 0.9429 | 0.9138 | 0.8855 | 0.8687 | 0.8606 | 0.8462 | 0.8268 | 0.8129 | 0.8068 | 0.7953 |
| 268 K | 4.741 | 4.961 | 5.186 | 5.412 | 5.614 | 5.810 | 6.007 | 6.175 | 6.275 | 6.348 | 6.517 |
|       | 0.9731 | 0.9638 | 0.9440 | 0.9187 | 0.9023 | 0.8850 | 0.8656 | 0.8559 | 0.8439 | 0.8425 | 0.8293 |
| 263 K | 4.730 | 4.959 | 5.179 | 5.383 | 5.617 | 5.801 | 5.984 | 6.166 | 6.271 | 6.332 | X |
|       | 0.9799 | 0.9652 | 0.9479 | 0.9339 | 0.9009 | 0.8895 | 0.8754 | 0.8596 | 0.8456 | 0.8491 | |
| 258 K | 4.726 | 4.951 | 5.183 | 5.393 | 5.603 | 5.792 | 5.976 | 6.146 | 6.244 | 6.331 | 6.498 |
|       | 0.9826 | 0.9699 | 0.9456 | 0.9287 | 0.9077 | 0.8935 | 0.8789 | 0.8682 | 0.8565 | 0.8495 | 0.8363 |
| 253 K | 4.718 | 4.946 | 5.175 | 5.381 | 5.577 | 5.765 | 5.973 | 6.124 | 6.224 | 6.313 | X |
|       | 0.9878 | 0.9729 | 0.9501 | 0.9349 | 0.9201 | 0.9059 | 0.8804 | 0.8775 | 0.8651 | 0.8568 | |
| 243 K | 4.723 | 4.924 | 5.151 | 5.371 | 5.558 | 5.759 | 5.948 | 6.117 | 6.212 | 6.297 | 6.444 |
|       | 0.9848 | 0.9857 | 0.9635 | 0.9403 | 0.9299 | 0.9090 | 0.8912 | 0.8805 | 0.8698 | 0.8631 | 0.8578 |
| 233 K | 4.713 | 4.928 | 5.137 | 5.355 | 5.545 | 5.742 | 5.933 | 6.095 | 6.189 | 6.259 | 6.429 |
|       | 0.9905 | 0.9835 | 0.9715 | 0.9487 | 0.9363 | 0.9171 | 0.8983 | 0.8901 | 0.8799 | 0.8788 | 0.8636 |
| 213 K | 4.704 | 4.905 | 5.115 | 5.312 | 5.511 | 5.688 | 5.886 | 6.049 | 6.132 | 6.220 | X |
|       | 0.9964 | 0.9975 | 0.9840 | 0.9717 | 0.9536 | 0.9432 | 0.9201 | 0.9105 | 0.9046 | 0.8956 | |
| 200 K | 4.702 | 4.892 | 5.096 | 5.300 | 5.485 | 5.674 | 5.851 | 6.015 | 6.112 | 6.188 | X |

|  | 0.9976 | 1.0053 | 0.9948 | 0.9782 | 0.9672 | 0.9502 | 0.9365 | 0.9260 | 0.9132 | 0.9096 |  |
|---|---|---|---|---|---|---|---|---|---|---|---|
| 193 K | X | X | X | X | 5.473<br>0.9739 | 5.670<br>0.9524 | 5.841<br>0.9415 | 6.010<br>0.9285 | 6.092<br>0.9223 | 6.172<br>0.9169 | X |
| 180 K | X | X | X | X | X | X | 5.826<br>0.9488 | X | X | X | X |
| 178 K | X | X | X | X | X | X | X | 5.974<br>0.9455 | 6.073<br>0.9309 | 6.151<br>0.9261 | X |
| 168 K | X | X | X | X | X | X | 5.819<br>0.9519 | 5.972<br>0.9463 | 6.067<br>0.9338 | 6.147<br>0.9278 | X |
| 165 K | X | X | X | X | X | X | X | X | X | X | 6.295<br>0.9199 |
| 158 K | X | X | X | X | X | X | X | 5.970<br>0.9471 | 6.051<br>0.9413 | 6.121<br>0.9400 | X |
| 153 K | X | X | X | X | X | X | X | X | 6.043<br>0.9451 | 6.098<br>0.9505 | X |
| 138 K | X | X | X | X | X | X | X | X | 6.029<br>0.9517 | 6.082<br>0.9581 | X |
| 133 K | X | X | X | X | X | X | X | X | X | X | 6.259<br>0.9358 |

**Table S3:** Steps of Molecular Dynamics simulation at each studied temperature. (After energy minimization the initial box was heated up to 340 K in order to avoid the aggregation of ethanol molecules. The stages of the simulation series for each composition were the following at each studied temperature (even at 340 K).)

| Type | Run time (ns) | Thermostat | time const. $\tau_T$ (ps) | Barostat | time const. $\tau_P$ (ps) |
|---|---|---|---|---|---|
| NPT_short | 2 | Berendsen[S1] | 0.1 | Berendsen | 0.1 |
| NPT_long | 10 | Nose-Hoover[S2-S3] | 1.0 | Parrinello-Rahman[S4] | 4.0 |
| NVT_short | 1 | Berendsen | 0.1 | | |
| NVT_long | 5 | Berendsen | 0.5 | | |

All results were derived from *NVT_long* simulations. In each case the TSSF-s were calculated also from the simulated models, using simulated partial radial distribution functions, by an in-house code. For calculating partial radial distribution functions the *g_rdf* software was used, that is part of the GROMACS software package.

Goodness-of-fit:

$R_W[F(Q)] = \sqrt{\frac{\Sigma_i(F^S(Q_i)-F^E(Q_i))}{\Sigma_i F^E(Q_i)}}$), where $F^S(Q)$ is the calculated value from the MD simulation, $F^E(Q)$ is the experimental structure factor.

---

(S[1]) Berendsen, H. J. C.; Postma, J. P. M.; DiNola, A.; Haak, J. R. Molecular dynamics with coupling to an external bath. *J. Chem. Phys.* **1984,** 81, 3684.
(S[2]) Nose, S. A molecular dynamics method for simulations in the canonical ensemble. *Mol. Phys.* **1984,** 52, 255-268.
(S[3]) Hoover, W. G. Canonical dynamics: Equilibrium phase-space distributions. *Phys. Rev. A* **1985,** 31, 1695.
(S[4]) Parrinello, M.; Rahman, A. Polymorphic transitions in single crystals: A new molecular dynamics method. *J. Appl. Phys.* **1981,** 52, 7182.

**Figure S2**: Selected partial radial distribution functions for the mixture with 40 mol % ethanol as a function of temperature.

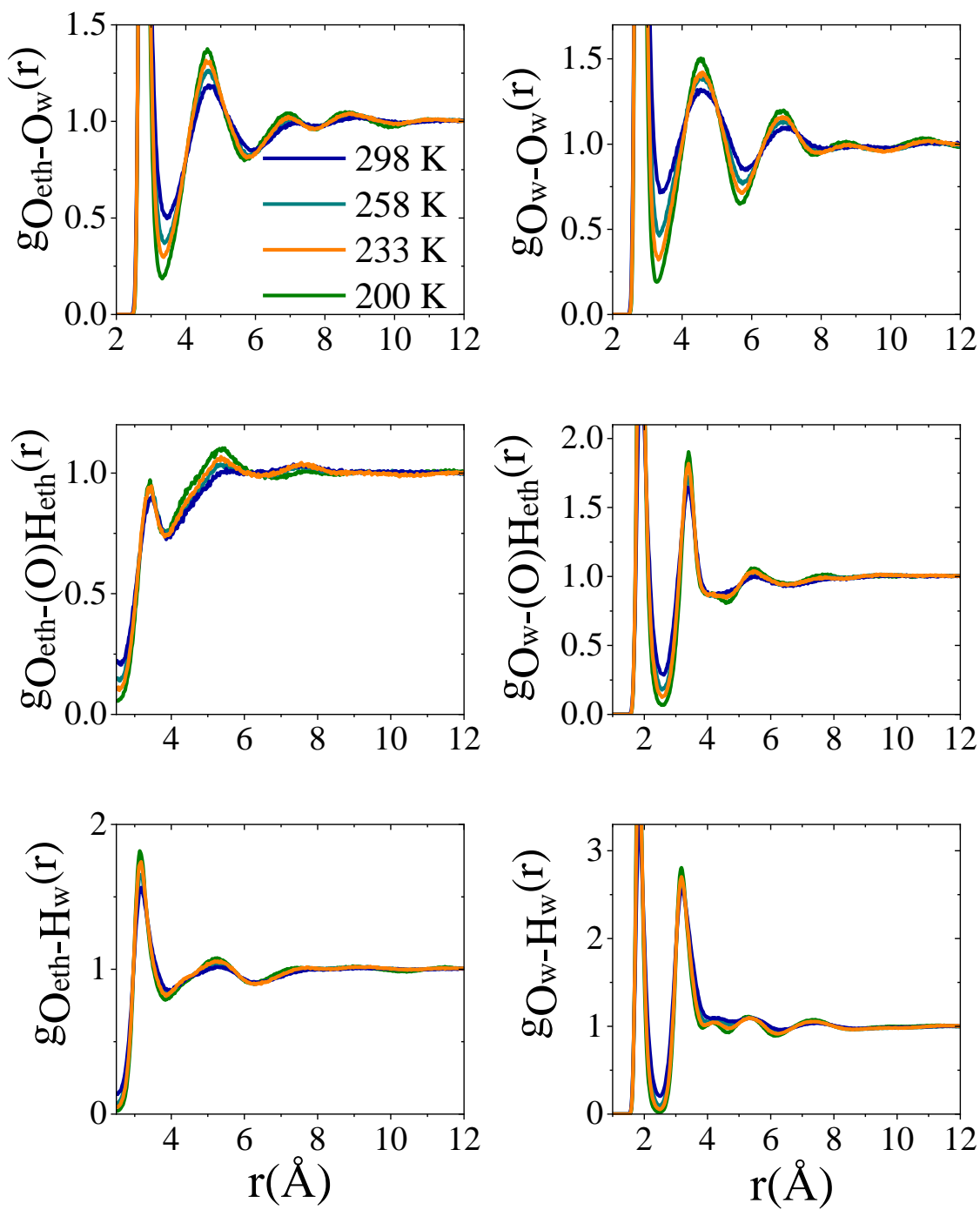

**Figure S3**: Selected partial radial distribution functions for the mixture with 50 mol % ethanol as a function of temperature.

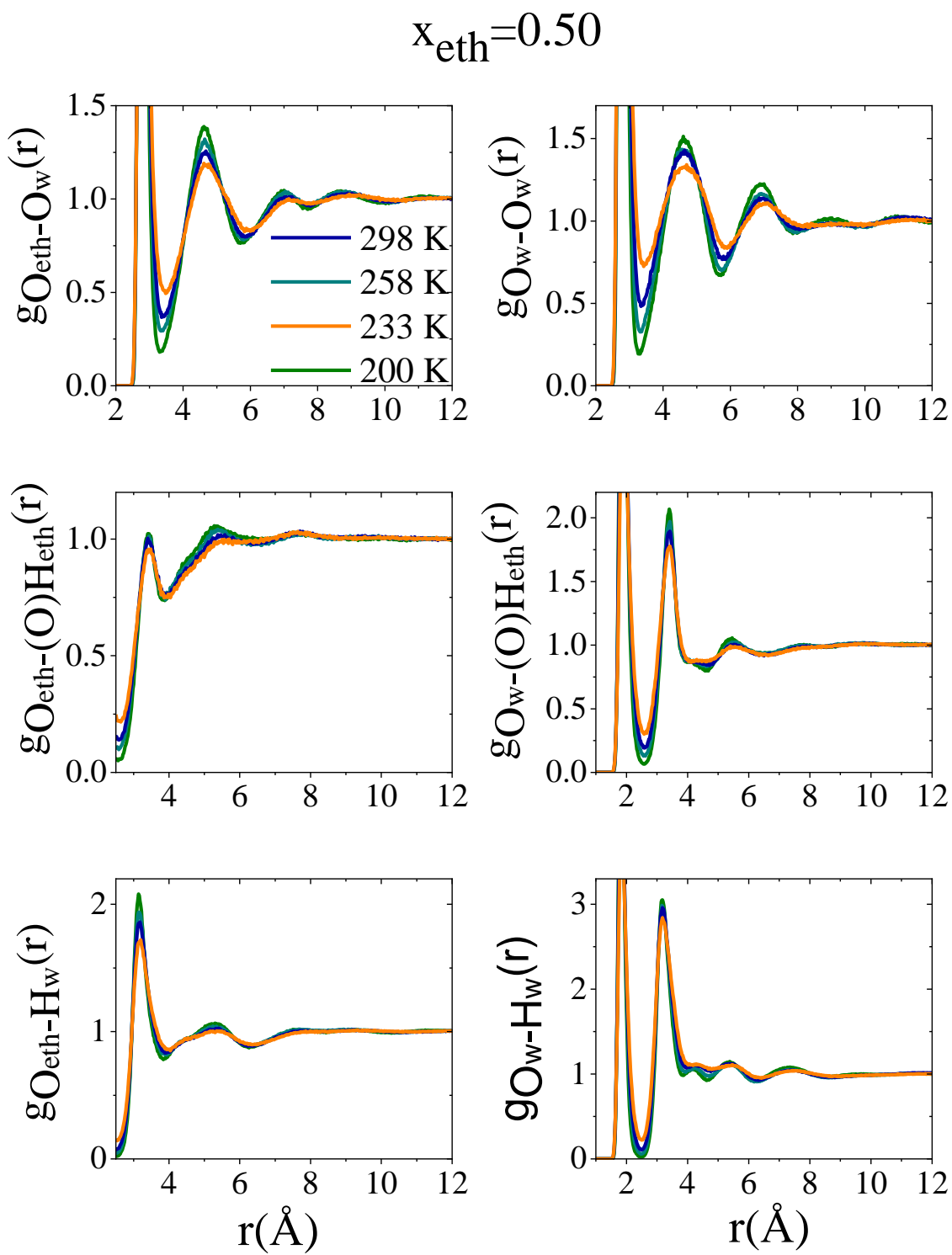

**Figure S4**: Selected partial radial distribution functions for the mixture with 60 mol % ethanol as a function of temperature.

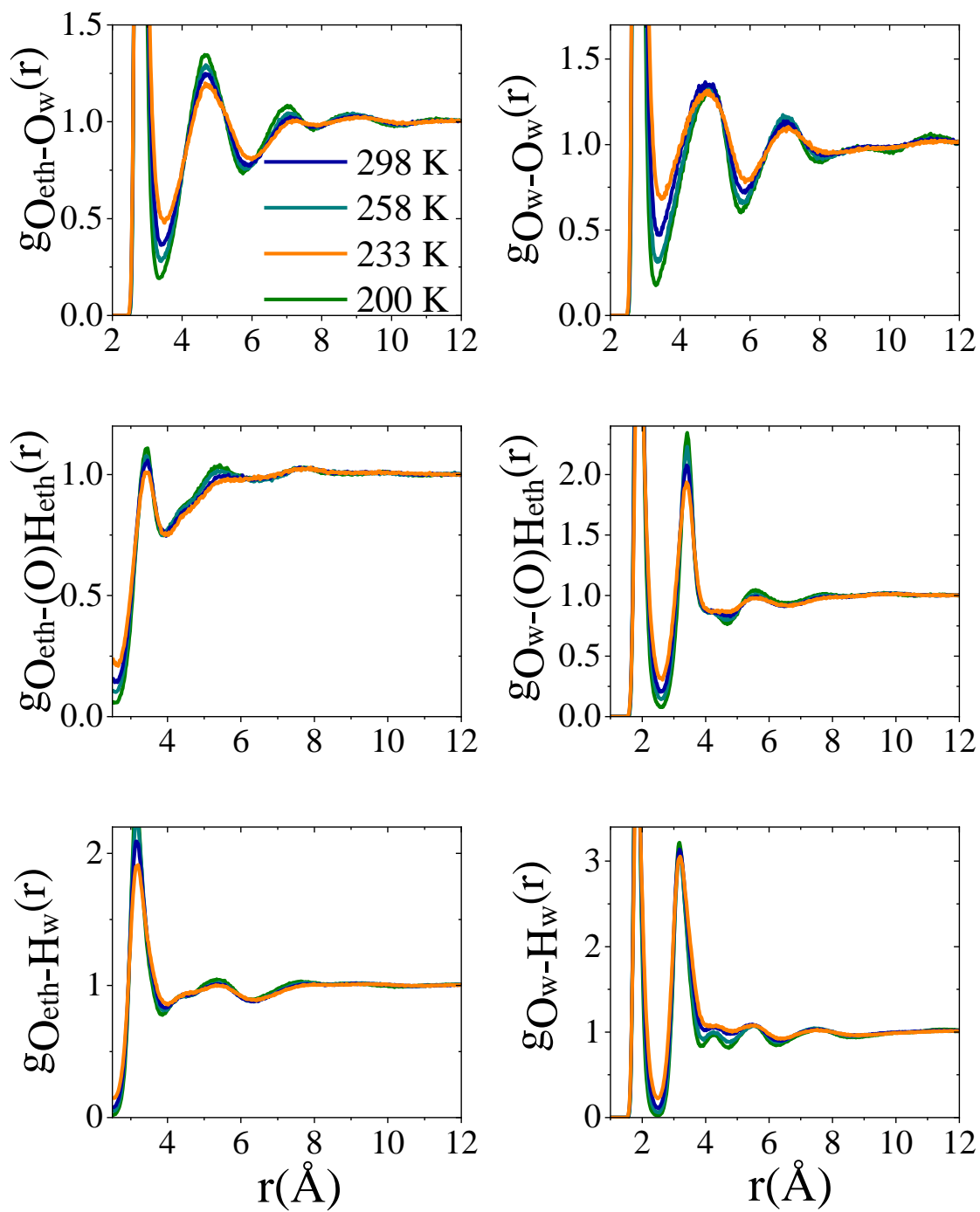

**Figure S5**: Selected partial radial distribution functions for the mixture with 70 mol % ethanol as a function of temperature.

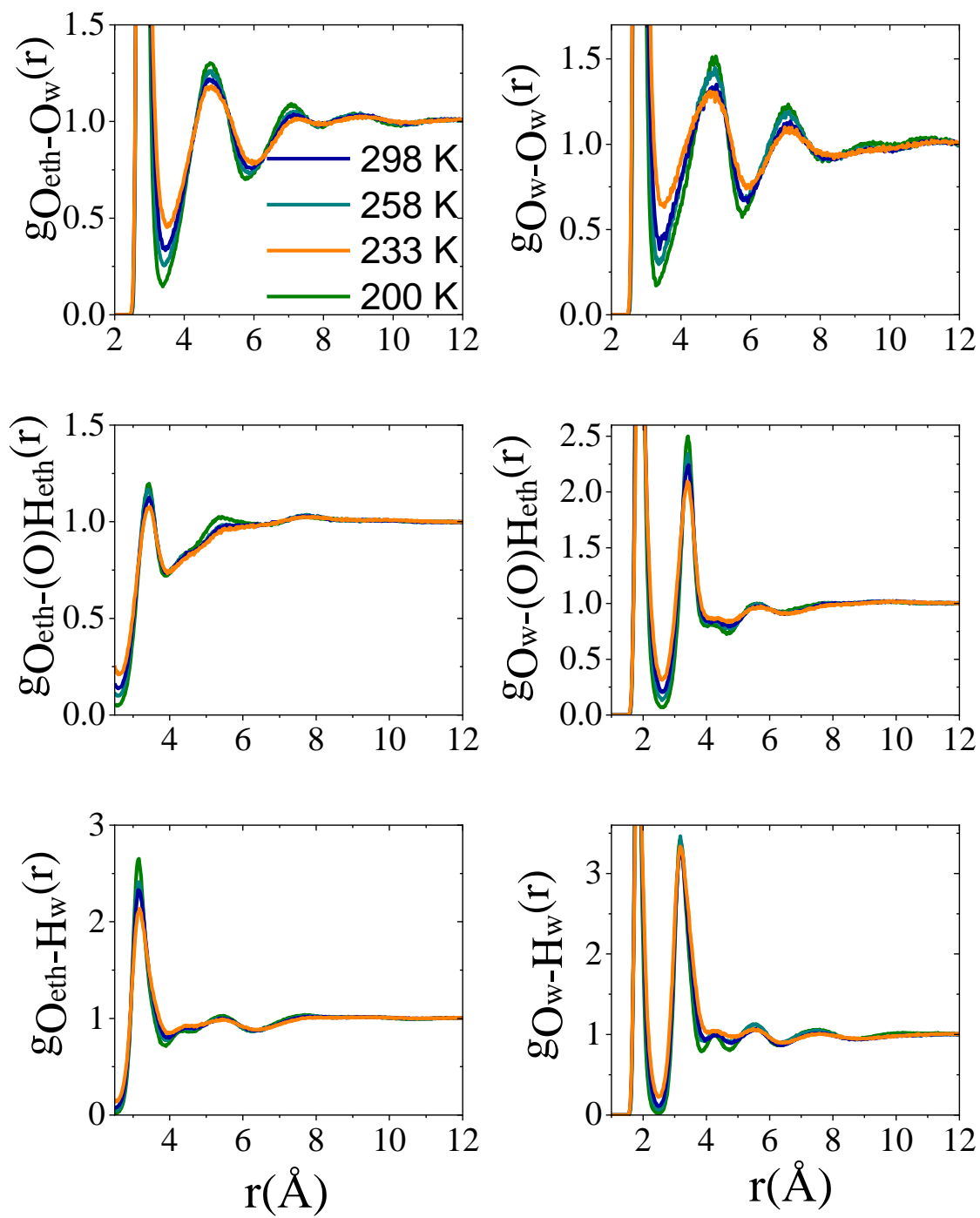

**Figure S6**: Selected partial radial distribution functions for the mixture with 80 mol % ethanol as a function of temperature.

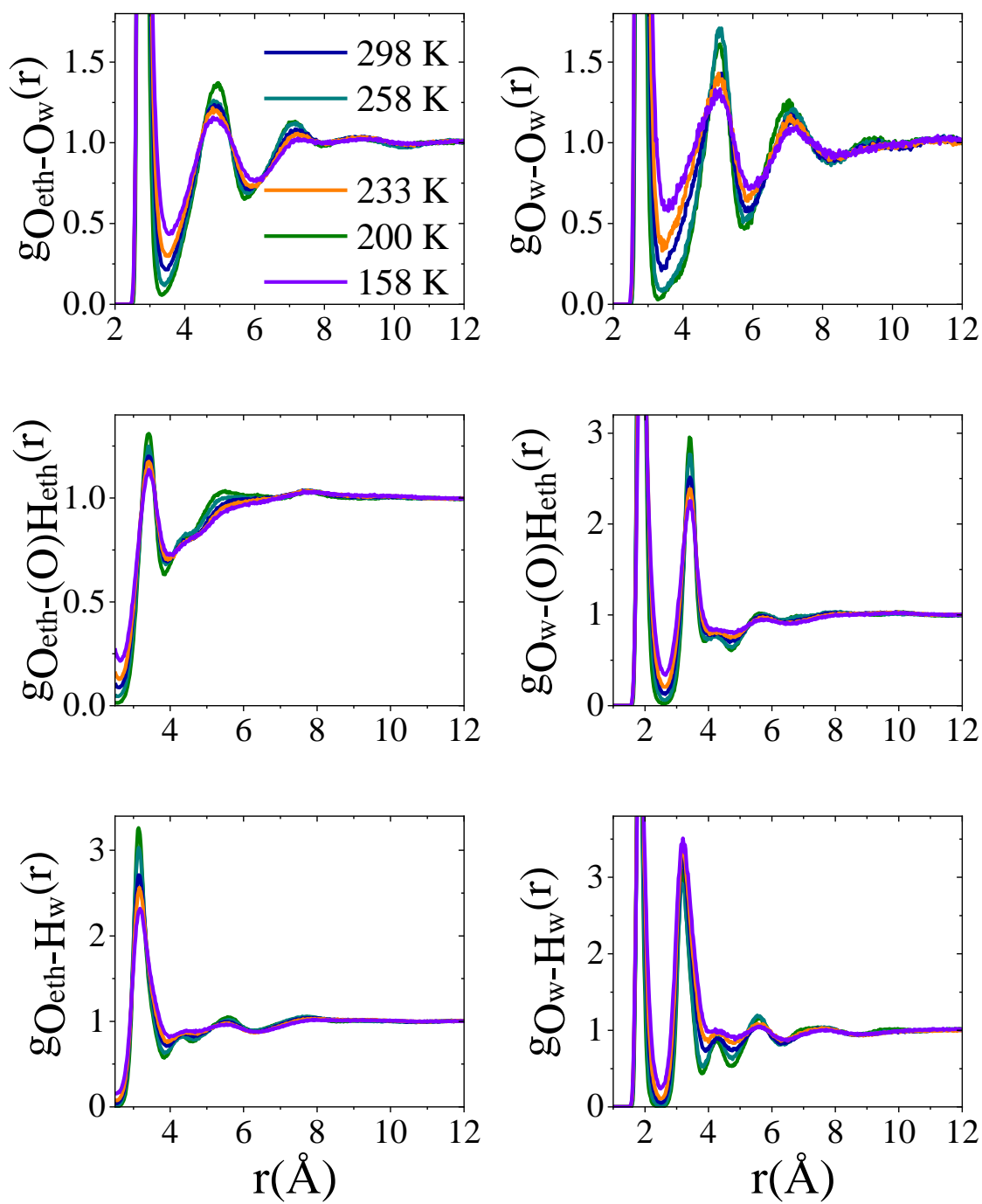

**Figure S7**: Selected partial radial distribution functions for the mixture with 85 mol % ethanol as a function of temperature.

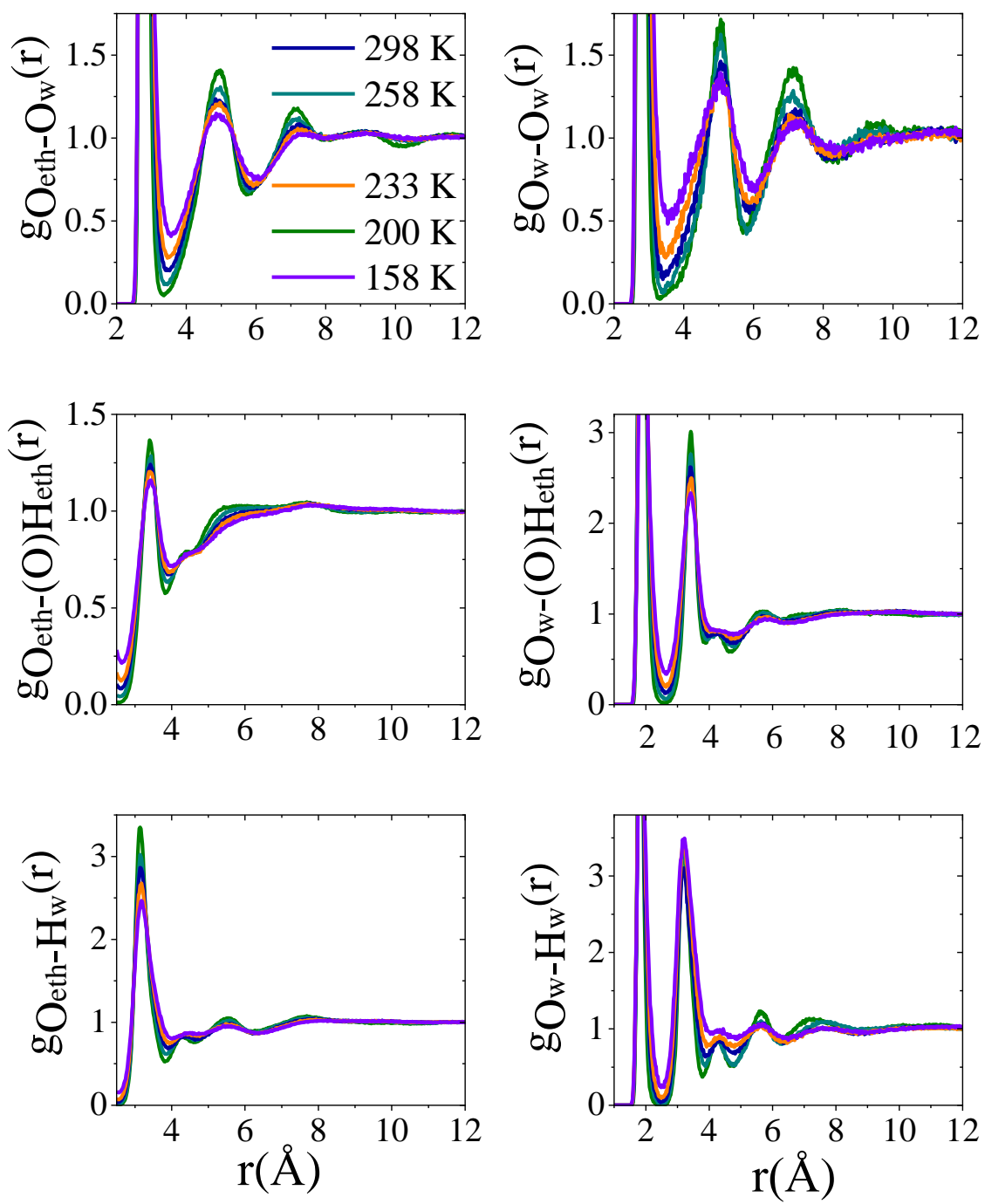

**Figure S8**: Selected partial radial distribution functions for the mixture with 90 mol % ethanol as a function of temperature.

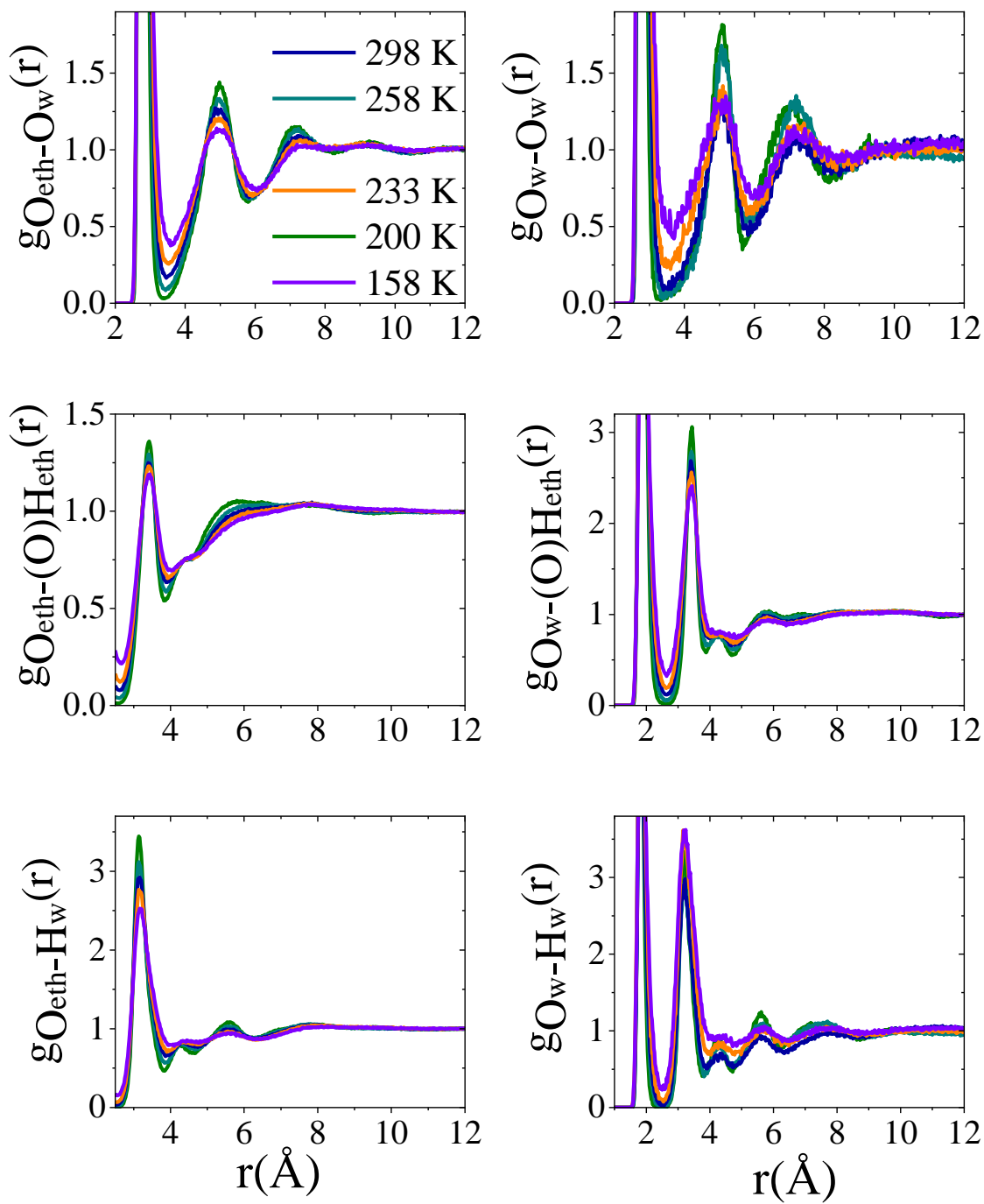

H-bond definition

Two molecules were considered hydrogen bonded (1) if they were found at a distance r(O···H) < 2.5 Å and H-O···O angle < 30° or (2) if they were found at a distance r(O···H) < 2.5 Å and the interaction energy between molecules was more negative than -3 kcal/mol (ca. -12 kJ/mol). In the latter case H-bond interactions are taken into account as attractive interactions. Note that results only with the "energetic" definition are presented in this work: the two definitions were in good agreement and the energetic definition, whenever it is available, is thought to be more robust.

The average number of hydrogen bonds ($n_{HB}$) in the mixtures (Figure S9), when taking into account all the connections, decreases when the ethanol content increases. At each concentration $n_{HB}$ linearly increases with decreasing temperature. Water subsystems follow this tendency, but only at ethanol concentrations lower than 60 mol%. At higher concentrations the number of H-bonds between water pairs is almost constant. This latter statement is, independently from the ethanol concentration, also true for the ethanol subsystem (Fig. S10) over the entire temperature range investigated.

**Figure S9.** Average H-bond numbers considering each molecule, regardless of their types, together with the case when considering water–water H-bonds only.

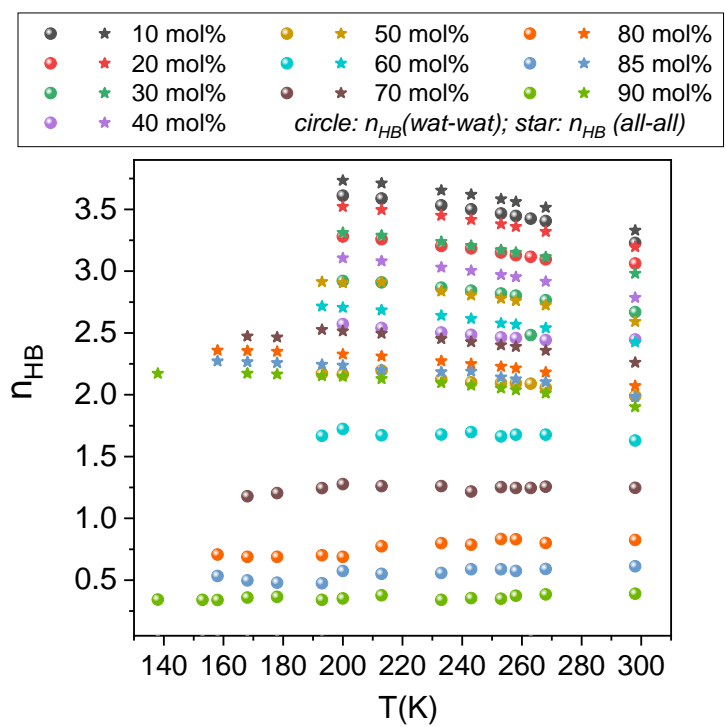

**Figure S10**: Average H-bond number for ethanol-ethanol subsystem.

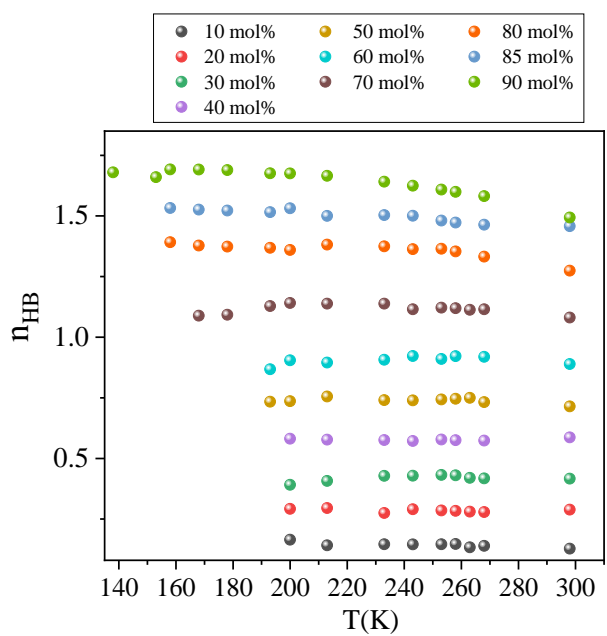

The number of H-bonded neighbors around (central) water and (central) ethanol molecules (Fig. S11) varies linearly with temperature at all concentrations. The only exception is the two highest ethanol concentrations (85 mol% and 90 mol%) below 190 K, where $n_{HB}$ becomes constant.

**Figure S11** Average H-bond numbers considering connections of water molecules only, as well as considering connections of ethanol molecules only.

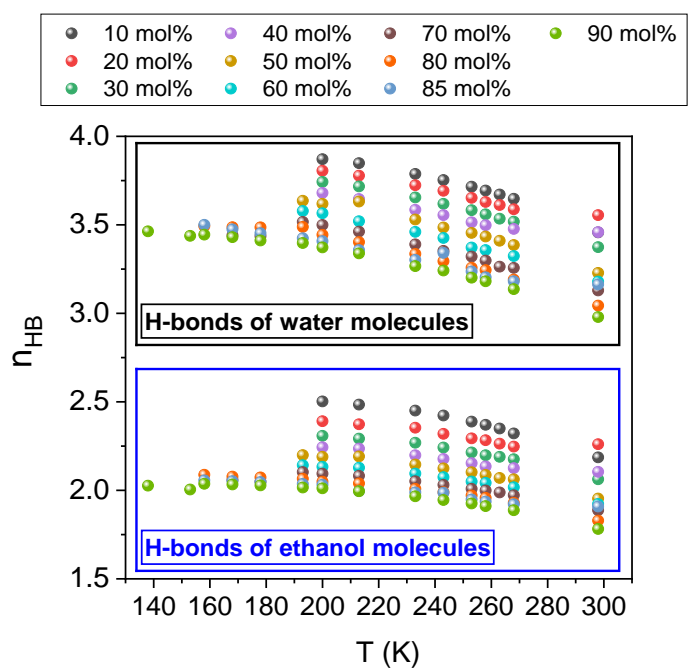

**Figure S12:** Fraction of donor and acceptor sites as a function of temperature: a) '1A:1D' for ethanol molecules; b) '1D:2A' for ethanol molecules; c) '2D:1A' for water molecules; d) '2D:2A' for water molecules.

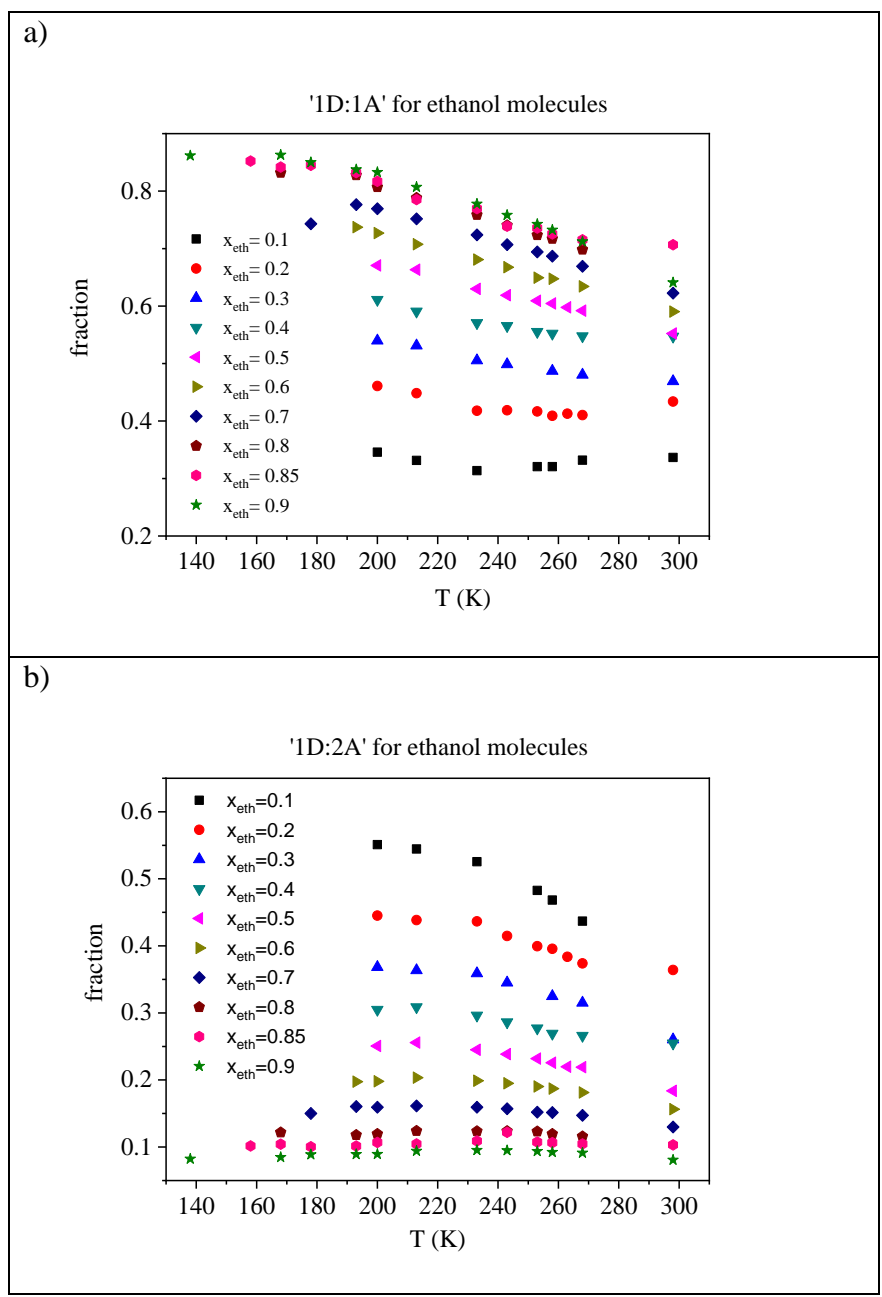

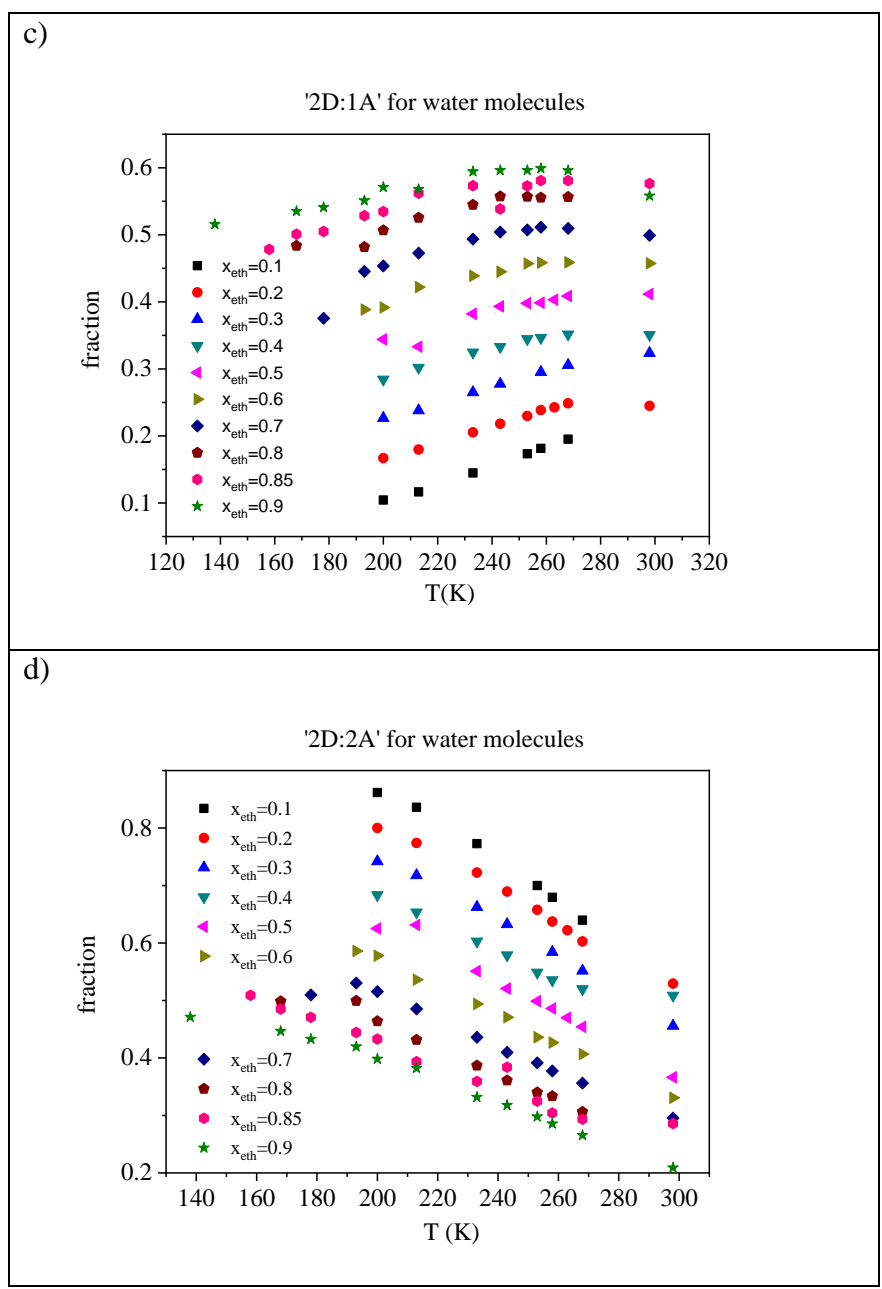

The behaviour of the '1A:1D' and '1D:2A' combinations for ethanol molecules, as well as that of the '2D:1A' and 2D:2A' combinations for water molecules, as a function of temperature, can be found in Fig. S12.

Concerning the '2D:2A' combination as temperature decreases the occurrence of this scheme case linearly increases for water molecules, at every concentration: this may be taken as an indication that '2D:2A' represents an inherent arrangement, being the most stable according to potential energy considerations.

H-bond number excess parameter

There are several approaches[S5-S7] for capturing deviations from characteristics of a system with randomly distributed molecules. The basic question is whether some type(s) of hydrogen bonds (e.g. water-water, water-ethanol, etc…) are preferred (i.e. more frequent than they are in a randomly distributed system). The following parameter is defined for characterizing a (possibly preferential) H-bonded environment around a central molecule:

$$f_{\alpha\beta} = \frac{n_{\alpha\beta}}{n_{\alpha all}} \frac{1}{x_\beta} \quad , \quad (1)$$

where $n_{\alpha\beta}$ and $n_{\alpha all}$ is the average H-bond number between $\alpha - \beta$ and $\alpha$ - all ($\alpha+\beta$) pairs, respectively. $x_\beta$ is the mole fraction of component $\beta$. In the case of an ideal (totally random) ethanol-water mixture this $f_{\alpha\beta}$ value is 1.0, whereas values higher than unity indicate preferential H-bonding.

Results are presented in Fig. S13. The $f_{wat-eth}$ and $f_{eth-eth}$ functions increase almost linearly with the increasing ethanol content. For $f_{eth-eth}$ a noticeable minimum can be detected at $x_{eth}=0.3$. On the other hand, in the case of $f_{wat-eth}$ a maximum emerges at $x_{eth}=0.6$. The $f_{wat-wat}$ function is above 1.0 over almost the entire concentration range. A well-defined maximum can be identified for $f_{wat-wat}$ around ethanol mole fractions 0.5-0.6 at 298 K, which is shifted at 200 K to ethanol mole fractions of 0.6-0.7. This corresponds to a significant excess of water molecules in the solvation shell of water. This maximum agrees well with the maximum of $G_{wat-wat}$ in Kirkwood-Buff integral theory.[S8-S10]

**Figure S13.** H-bond number excess parameter: parameter for characterizing the preferential H-bonded environment around a central molecule ($f_{\alpha\beta}$). Black solid circle symbols: $f_{wat-wat}$ at 298 K; red open circle symbols: $f_{wat-wat}$ at 200 K; black solid triangle symbols: $f_{wat-eth}$ at 298 K; green open triangle symbols: $f_{wat-eth}$ at 200 K; black solid square symbols: $f_{eth-eth}$ at 298 K; blue open square symbols: $f_{eth-eth}$ at 200 K.

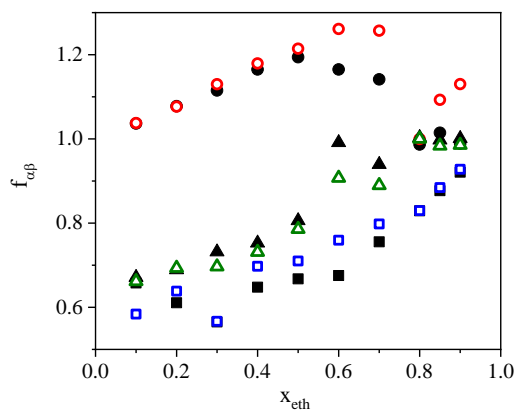

**Figure S14.** The average largest cluster size (C1) and average second largest cluster size (C2) as a function of ethanol concentration and temperatures in ethanol-water mixtures. Violet sphere symbols: $x_{eth}=0.1$; orange sphere symbols: $x_{eth}=0.2$; navy sphere symbols: $x_{eth}=0.3$; dark cyan sphere symbols: $x_{eth}=0.4$; magenta sphere symbols: $x_{eth}=0.5$; black sphere symbols: $x_{eth}=0.6$; green sphere symbols: $x_{eth}=0.7$; red sphere symbols: $x_{eth}=0.8$; blue sphere symbols: $x_{eth}=0.85$; dark red sphere symbols: $x_{eth}=0.9$.

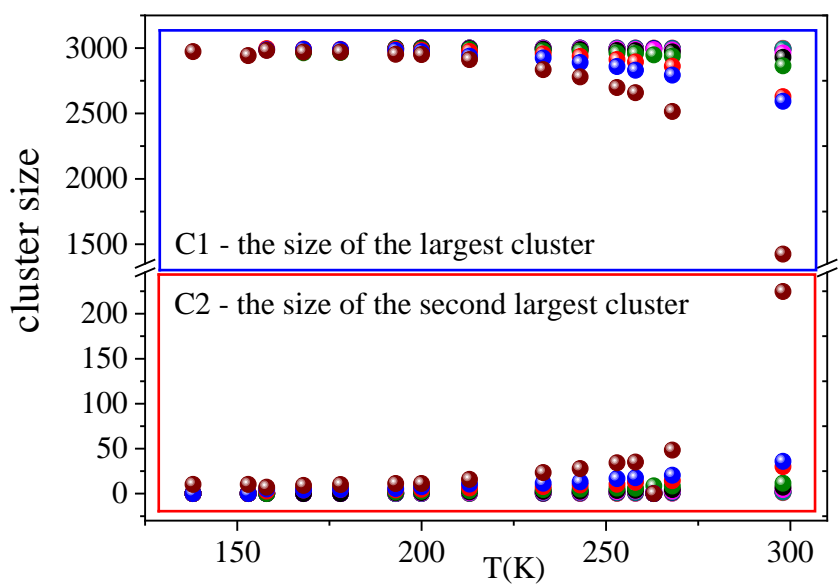

The average largest cluster size (C1) and average second largest cluster size (C2) are plotted in Fig. S14. Several works[S11-S14] have already proved that the properties of these quantities are good indicators for determining the location (concentration-temperature pair) of the percolation transition. At low temperatures, below 230 K, the number of molecules in the largest cluster equals the total number of molecules in the systems. It also means that the value of C2 is almost zero at each studied concentration. In these cases, molecules percolate throughout the systems. Around 230 K the spheres corresponding to the different concentrations start to move away from each other, i.e., the largest clusters start to shrink. The largest cluster at $x_{eth}=0.9$ and 300 K is only half the size of what it was at 150 K, which means that percolation is questionable. The average second largest cluster size shows that the system contains several smaller assemblies of less than 200 molecules. Note that the only case where the question of existence of percolation threshold arises is the 90 mol% solution at room temperature.

Fractal dimension of the largest cluster:

According to random site percolation theory, infinite clusters are true fractals at the percolation threshold with fractal dimension $f_d=2.53$ in three dimensions, and $f_d=1.896$ in two dimensions.[S11-S13] It has already been shown[S11-S13] that we cannot detect a percolated cluster with $f_d$ value smaller than 2.53 in three, and 1.896 in two dimensions.

In the light of the foregoing, among the systems studied the one with 90 mol% ethanol deserves further scrutinization from the point of view of the percolation threshold. It was found that the *fractal dimension of the largest cluster* is 2.73 at 298 and 2.90 at 200 K, respectively. These values are significantly larger than the corresponding values to the percolation threshold. It can be stated that the largest cluster forms a 3D percolated network still at $x_{eth}=0.9$. For the other systems studied, $f_d$ values were larger than 2.8, which confirms that their largest clusters form 3D percolated networks.

**Figure S15**. Cluster size distributions from the room temperature to the lowest studied temperature a) for $x_{eth}$=0.85, b) for pure ethanol.

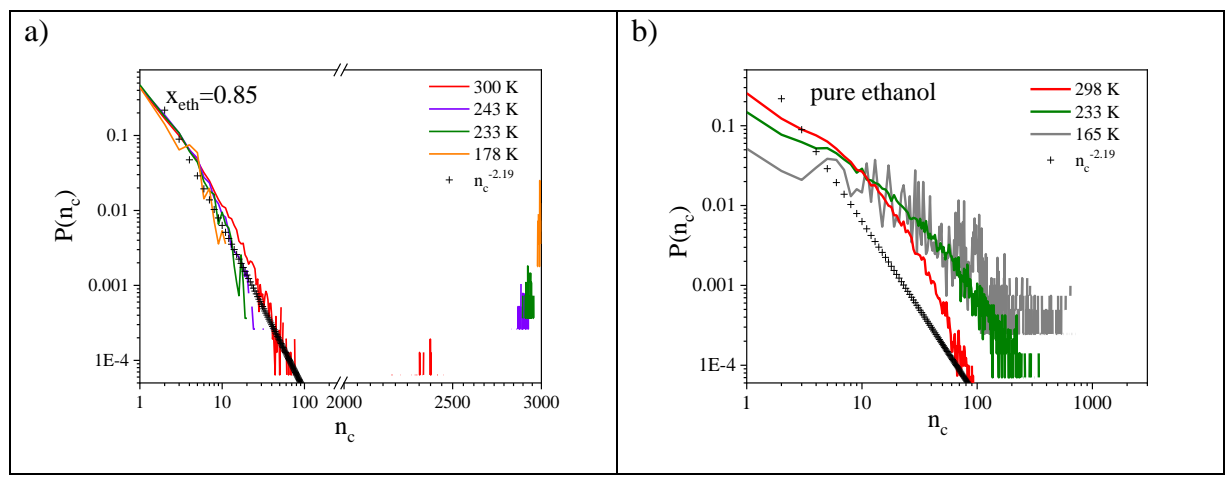

**Figure S16.** Typical hydrogen bonded network topologies in water-ethanol mixtures at concentrations $x_{eth}$ = 0.40 (left), 0.70 (middle), and 0.90 (right). Red symbols indicate the molecules and green symbols represent the H-bonds.

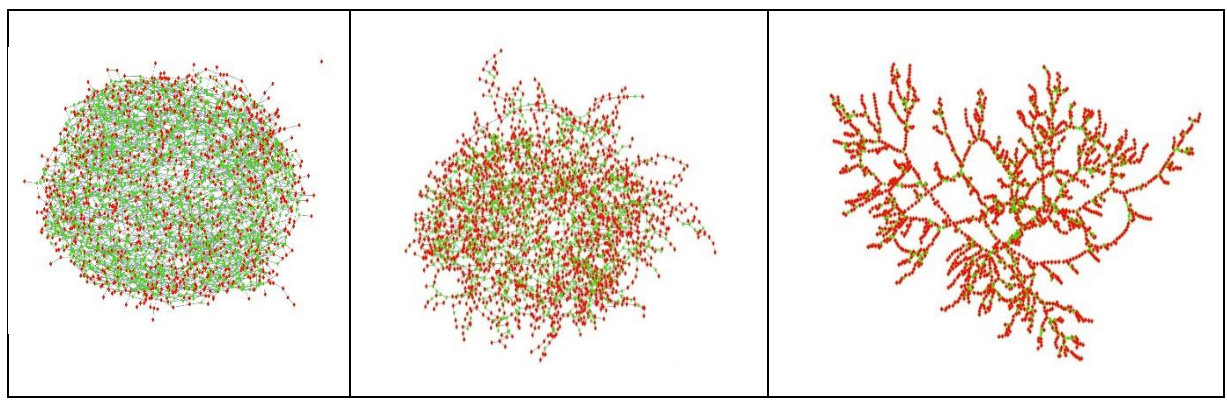

At low ethanol concentrations, these systems contain mainly cyclic entities, whereas at higher alcohol concentrations more and more chains appear. For $x_{eth}$=0.9, even a dendrite type structure can be found that exists, for instance, in the neural networks.[S15,S16]

Background of the Laplace spectra

The structure of a network can be fully characterized with the adjacency or the combinatorial Laplace (L) matrices.[S17-S28] The Laplace matrix can be defined as follows:

$$L_{ij} = k_i \delta_{ij} - A_{ij} \quad (1)$$

where $k_i$ is the number of (hydrogen) bonded neighbours of molecule 'i'; $\delta_{ij}$ is the Kronecker delta function and $A_{ij}=1$ if a bond exists between nodes i and j.

It is known that the Laplacian matrix is positive semidefinite and has nonnegative eigenvalues.[S17,S18] Furthermore, 0 is always an eigenvalue of L and the multiplicity of the eigenvalue 0 is equal to the number of the connected components of the graph.

'Stability' of H-bonded networks

Finally, we would like to provide some indicator for the 'stability' of H-bonded networks found in ethanol-water mixtures, as a function of temperature. Below we wish to devise a simple number that is related of the number of H-bonds that needed to be removed so that the network in question would not be percolating any longer.

The most important theorem is coupling to the connectivity and to the second smallest positive eigenvalue (Fiedler eigenvalue) of a Laplacian is known as the Cheeger inequality[S29, S30]:

$$\frac{\lambda_2}{2} < h(G) < \sqrt{2\lambda_2} \qquad (2)$$

where h(G) is the Cheeger constant (or conductance) of a graph G. This inequality is related to the minimum number of bonds such that, when removed, cause the graph to become disconnected ('non-percolated' according to the terminology of H-bonded networks) [S31]. Therefore, h(G) (or a similarly derived quantity like $\lambda_2 * n_{HB}$, see below) can serve as a well-defined parameter to measure the 'distance' from the percolation transition. Some applications of this theory for molecular liquid can be found in Ref. S32. and S33-S35.

The h(G) quantity is defined by the following equation.

$$h(G) = \min \frac{E(Sc, V-Sc)}{\min(Vol(Sc), Vol(V-Sc))} \qquad (3)$$

Here, V–$S_C$ and $S_C$ are two non-empty subsets of V nodes of the G graph, Vol($S_C$) and Vol(V–$S_C$) are the sums of the number of both intra- and inter-set connections of each node of the given subset. The E(V–$S_C$, $S_C$) is the number of inter-set links, connect nodes belonging to the different subsets. This expression has a minimum if the number of inter-set links (i.e. bonds) is taken as 1 and the denominator is a value indicating the half of the "volume" size of the entire system (Vol(V)/2, equals to the number of links of the G graph or by other words: the total number of hydrogen bonds in the configuration). We can call this value $h(G)_{min}$. After simple mathematical transformations, we can obtain the following inequality from Eq. 2.:

$$\frac{\lambda 2}{2h(G)min} \leq \frac{h(G)}{h(G)min} \leq \frac{\sqrt{2\lambda 2}}{h(G)min} \qquad (4)$$

The left and the right sides of the inequality defined by Equation 4 are shown in Figure S17. This inequality is related to the minimum number of bonds that, when removed, cause the graph to become disconnected ('non-percolated'). This inequality provides a lower and upper limit on the stability of the percolated network, considering also the effect of finite size. It can be seen that the stability of the hydrogen bond network decreases significantly with increasing ethanol concentration.

**Figure S17.** Values of the inequality calculated by Equation 4. Black open squares: left side of Eq. 4 at 298 K; black solid squares: right side of Eq. 4 at 298 K; red open circles: left side of Eq. 4 at 233 K; red solid circles: right side of Eq. 4. at 233 K.

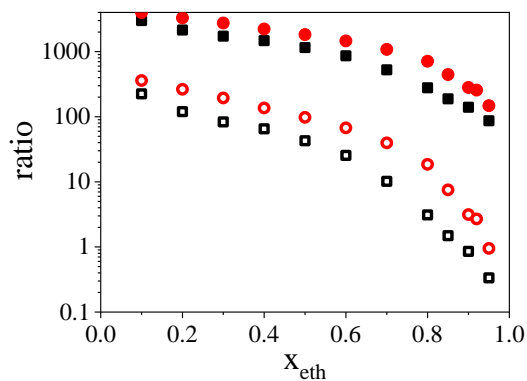